\documentclass[a4paper,10pt]{article}
\usepackage{
times,
amsmath,amssymb,epsfig,graphics,graphicx}

\newcommand{\sech}{\textrm{ sech }}

\newcommand{\be}{\begin{eqnarray}}
\newcommand{\ee}{\end{eqnarray}}

\begin{document}

\begin{center}
{\Large\bf Rotational Symmetry and Degeneracy: A Cotangent-Perturbed Rigid 
Rotator of Unperturbed Level Multiplicity}
\end{center}
\vspace{0.02cm}

\begin{center}
D.\ E.\ Alvarez-Castillo$^1$, C.\ B.\ Compean$^2$, and M.\ Kirchbach$^3$
\end{center}

\vspace{0.01cm}
\begin{center}
$^1$ {\it H.Niewodnicza\'nski Institute of Nuclear Physics, 
Radzikowskiego 152, 31-342 Krak\'ow, Poland}\\
$^2$ {\it Instituto de F\'{\i}sica y Matematica},\\
{\it Universidad Michoacana de San Nicol\'as de Hidalgo,
Edificio C-3,}\\
{\it Ciudad Universitaria, Morelia, Michoac\'an 58040, M\'exico}\\
$^3${\it Instituto de F\'{\i}sica},
         {\it Universidad Aut\'onoma de San Luis Potos\'{\i}},\\
         {\it Av. Manuel Nava 6, San Luis Potos\'{\i}, S.L.P. 78290, M\'exico}
\end{center}

\vspace{0.01cm}

\begin{flushleft}
{\bf Abstract:} 
We predict level degeneracy of the rotational type 
in diatomic molecules 
described by means of a  cotangent-hindered rigid rotator.
The problem is shown to be exactly solvable
in terms of  non-classical Romanovski polynomials.
The energies of such a system  are  linear combinations 
of $t(t+1)$ and $1/\lbrack t(t+1)+1/4 \rbrack $ terms with the 
non-negative integer principal quantum number 
$t=n+|{\widetilde m}|$  being the sum of the order $n$ of the 
polynomials and the  absolute value, $|{\widetilde m}|$, 
of the square root of the separation constant between the polar and 
azimuthal angular motions.
The latter obeys with respect to $t$
same branching rule, $|{\widetilde m}|=0, 1,..., t$, as does the
magnetic quantum number with respect to the
 angular momentum, $l$,  and in this fashion
 the $t$  quantum number presents itself formally  
indistinguishable  from $l$.  In effect, the spectrum of the hindered rotator
has the same $(2t+1)$-fold level multiplicity as the unperturbed one. 
For low $t$ values  wave functions and excitation energies 
of the perturbed rotator  differ from the ordinary spherical harmonics,
and the $l(l+1)$ law, respectively,  while
approaching them asymptotically with the increase of $t$.
In this fashion the breaking of the rotational symmetry at the 
level of the representation functions is opaqued by the level degeneracies.
The model furthermore provides a tool for the description of rotational bands
with anomalously large gaps between the ground state and its 
first excitation.

\end{flushleft}

\section{Relationship between symmetry and degeneracy:Introductory remarks.}
The relationship between the rotational symmetry of the central potentials
and the degeneracy in their spectra  is  a  complex one.  
One expects that due to rotational symmetry
the spectra of the central potentials will show the
$(2l+1)$-fold degeneracy with respect to the magnetic quantum number, 
which they really do, though not alone. Matters are complicated 
by the circumstance that various exactly solvable
radial potentials (see ref.~\cite{Dutt} for a compilation)
have more symmetries than the rotational symmetry of the angular motion.
The reason for this is that in the process of the reduction of the radial part
of the separated  Schr\"odinger 
equation down to the hyper-geometric differential equation by an 
appropriate point-canonical transformation, the energy always
emerges as a combination, known as principal quantum number,
of $l$, and the order $n$ of the polynomial.
In many cases the radial wave equation can be transformed to a 
power series of the Casimir operator of some Lie group different 
from SO(3) and the principal quantum number can be associated with 
the corresponding eigenvalues of the group invariant. 
A prominent example in that regard 
is the fundamental inverse distance problem, 
where this combination appears as, $N=l+n+1$,  
the positive integer $N$ being known as the principal quantum number.
Summing up the $(2l+1)$ fold degeneracies for all $l=0,1,2,..(N-1)$, 
a larger  $\sum_0^{N-1}(2l+1)=N^2$-fold  level degeneracy occurs 
with respect to both $l$ and $m$. The
rotational $(2l+1)$-fold multiplicity is of course  at the very root of the 
larger degeneracy. In the particular case of the Coulombic potential
 it is possible to associate the  larger degeneracies
with the larger symmetry group $SO(4)$ due to the circumstance  that
the respective momentum space wave functions can be identified
with the hyper-spherical harmonics on the three-dimensional spherical
surface, $S^3$ \cite{Kim} whose isometry group is  $SO(4)$. Then,  
$N-1=(n+l)$
acquires meaning of  four-dimensional angular momentum. 
Within this context, the $(2l+1)$ fold degeneracy is manifest at a lower
level of the group reduction chain, $SO(4)\supset SO(3)\supset SO(2)$.

It is commonly accepted that in order to reveal the beyond-rotational 
symmetry of a central potential problem one has to find a 
suited surface of a given isometry group, ${\mathcal G}$,
such that, upon an appropriate transformation,
the radial Schr\"odinger wave functions can be mapped onto
the bases of the irreducible representations of ${\mathcal G}$
in which case the former would describe 
free geodesic motion on the ${\mathcal G}$ manifold. 
For example, the radial Schr\"odinger equations 
with the $\csc^2\left(\frac{r}{R} \right)$, and
$\sech^2\left( \frac{r}{d}\right)$ potentials, where $R$ and $d$ 
stand for some matching length parameters which make the 
arguments dimensionless, can be transformed in their turn
to the problems of the four dimensional rigid rotator on the three-dimensional
spherical surface, $S^3$, and the 
hyperbolic rotator on the two-dimensional one sheeted hyperboloid,
${\mathbf H}^{1+1}_1$ \cite{Rocha}. As long as the isometry
groups of the $S^3$- and the ${\mathbf H}^{1+1}_1$ surfaces are
$SO(4)$, and $SO(1,2)$ respectively, 
the $\csc^2$ and $\sech^2$ potentials are said to
be $SO(4)$ and $SO(1,2)$ symmetric. 
Stated differently, the radial Schr\"odinger equations under 
investigation can be  transformed to angular, better, rotator,  problems on
$S^3$ and ${\bf H}^{1+1}_1$ thus making their respective beyond-rotational 
symmetries manifest. 

The above considerations teach one two lessons. The first is that in general
complete Schr\"odinger equations
are not well suited as templates for studying the relationship
between rotational degeneracies and symmetry because
the rotational $(2l+1)$-fold multiplicity of the states
may appear surmounted  by the possible additional symmetries 
enjoyed by those potentials.  

The second is that the best template for studying  
any symmetry  is a ``rotator'', i.e. a free geodesic motion on
a surface whose isometry group coincides with the 
symmetry under investigation.

We here are specifically interested in the relationship between 
rotational symmetry and degeneracy and shall focus attention on
the rigid rotator on the two-dimensional spherical surface $S^2$
whose isometry group is $SO(3)$.
It is described by the angular part of the Laplace 
operator and is known to equal (up to $-1/r^2$) the 
squared orbital angular momentum. 
 For a constant radius, $r=a$, 
\begin{eqnarray}
\nabla^2= -\frac{1}{a^2}{\mathbf L}^2, \quad r=a=\mbox{const},
\end{eqnarray}
the Schr\"odinger problem  reduces to that of the
rigid rotator.  
The quantum mechanical Hamiltonian for the free three dimensional
rigid  rotator is then the angular part of the Schr\"odinger equation
and given by
\begin{equation}
\frac{\hbar ^2}{2{\mathcal I}}{\mathbf L}^2Y _{l}^{m}(\theta, \varphi)=
\frac{\hbar^2}{2{\mathcal I}}l(l+1)Y_l^m(\theta, \varphi), 
\quad {\mathcal I}=\mu a^2,  
\label{I1}
\end{equation}
where now $a$ denotes the radial distance between the two bodies 
of the rotator,
$\mu$ stands for their reduced mass, ${\mathcal I}$ denotes their 
inertial moment, and $Y_l^m(\theta ,\varphi)$ are the 
${\mathbf L}^2$ eigenfunctions. These are the standard spherical harmonics,
which constitute the bases of the $(2l+1)\times (2l+1)$ dimensional
irreducible $SO(3)$ representations, a reason
for the observed  $(2l+1)$--fold level degeneracy \cite{AW}.
In the following we shall see that the rigid rotator problem provides
an adequate template for studying  the relationship between degeneracy and 
symmetry. 

To be specific, we shall perturb the rigid rotator by a cotangent function
of the polar angle in which case the associated Legendre equation can be
transformed to an one-dimensional Schr\"odinger equation with the 
trigonometric Rosen-Morse potential. This potential is exactly solvable 
and  one  can write down  exact energies and wave functions of the 
hindered rotator. The cotangent-perturbed rigid rotator
has been mentioned independently 
in refs.~\cite{gangopadh}, and \cite{Levai}. 
Both references entertain the possibility of introducing 
the trigonometric Rosen-Morse potential as a non-central potential
in the Schr\"odinger equation with a Coulombic potential without really 
executing the calculation. In a work \cite{David} prior to this,
the cotangent interaction has been introduced as a non-central potential in
the polar part of the full Schr\"odinger problem and handled 
along the line of ref.~\cite{gangopadh}.
According to the prescription suggested in
ref.~\cite{gangopadh} the angular equation under investigation
has been subjected to a somewhat artificial
transformation toward the hyperbolic Scarf potential
with the aim to study possible $SO(2,1)$ symmetry properties of the
latter. We here instead  take the straight path and solve the equation
for the associated Legendre functions in the presence of a 
cotangent perturbance directly. 
In so doing we find a result which we consider interesting to be reported on 
in so far as it sheds some surprising light  on the relationship 
between degeneracy and symmetry. Namely, we find that the rotational
$ (2l+1)$-fold level degeneracy of the unperturbed rotator is respected by
the hindered one.\\
 The paper is organized as follows.
In the next section we briefly review the rigid rotator problem 
for the sake of self-sufficiency of the presentation and fixing notations.
Section 3 is devoted to the problem of the hindrance of the
rigid rotator by the cotangent interaction. There, we present
the exact solutions, and discuss the phenomenon 
of observing rotational  $(2l+1)$-fold level degeneracy 
despite perturbance of the rotational invariance.
We shall refer to this phenomenon  as a ``symmetry breaking opaqued by 
level degeneracies'',  or, shortly,  ``opaqued symmetry breaking''. 
The paper closes with brief conclusions.

\section{The quantum rigid rotator}

The explicit expression for {\bf L}$^2$ reads,
\begin{equation}
{\mathbf L}^2=-
\left[ 
\frac{1}{\sin \theta }\frac{
\partial }{\partial \theta}
\sin \theta \frac{\partial }{\partial \theta}
-\frac{1}{\sin^2\theta }\frac{\partial^2}{\partial \varphi^2}
\right], 
\label{I2}
\end{equation}
with $\theta \in [0,\pi]$, and $\varphi \in [0,2\pi )$.
Separating variables in eq.~(\ref{I1}) amounts to
\begin{eqnarray}
-\left[ 
\frac{1}{\sin \theta }\frac{
{\mathrm d} }{{\mathrm d}\theta}
\sin \theta \frac{{\mathrm d}}{{\mathrm d}\theta}
-\frac{m^2}{\sin^2\theta }
\right] P_l^{|m|}(\theta) &=&l(l+1)P_l^{|m|} (\theta),
\label{Sch_pol}\\
\frac{\partial ^2\Phi (\varphi) }{\partial \varphi ^2}=
-m^2 \Phi (\varphi), &\quad& \Phi (\varphi)=e^{im\varphi},
\label{Gl1}
\end{eqnarray}
where $P_l^{|m|}(\theta )$ are the standard associated 
Legendre functions \cite{AW}. The resulting ${\mathbf L}^2$ eigenfunctions
are,
\begin{equation}
P_l^{|m|}(\cos \theta )e^{im\varphi }=Y_l^m(\theta, \varphi),
\end{equation}
i.e. the standard spherical harmonics, $Y_l^m(\theta ,\varphi)$,
which thereby define the basis of
the $(2l+1)\times (2l+1)$ dimensional irreducible representation of the
rotational group in three dimensions, $SO(3)$.
In view of the fact that eqs.~(\ref{Gl1}) describe the free geodesic motion on
$S^2$, the standard spherical harmonics will occasionally be termed to as
wave functions of the free geodesic motion on $S^2$.

The rigid rotator is commonly used as a tool in the description of
diatomic molecules \cite{Gerrit}--\cite{Hathorn}, a field that has recently 
enjoyed a significant boost  through the progress in 
ultra-cooling  techniques. 
The introduction of a potential in eq.~(\ref{I1}) has the effect to
subject the rotator to an interaction, a procedure  often referred to
as  perturbance, or hindrance of the  rotator.  Numerically solvable 
angular potentials like
$V(\theta )=V_0\cos^2\theta$, and $V(\theta )=\frac{V_0}{2}(1-\cos^2\theta )$
have been introduced in refs.~\cite{Curl}, and \cite{MacRury}, respectively,
and subsequently employed  by several authors.

We here shall consider a rigid rotator hindered by a cotangent function of the
polar angle in which case the first eq.~(\ref{I1}) extends toward,
\begin{eqnarray}
-\left[ 
\frac{1}{\sin \theta }\frac{
{\mathrm d} }{{\mathrm d}\theta}
\sin \theta \frac{{\mathrm d}}{{\mathrm d}\theta}
-\frac{{\widetilde m}^2}{\sin^2\theta }
\right]F (\theta) -2b\cot \theta F(\theta ) &=& 
\epsilon  F(\theta),
\label{master}\\
-\frac{\partial^2}{\partial\varphi ^2 }e^{i{\widetilde m}\varphi}&=&
{\widetilde m}^2e^{i{\widetilde m}\varphi}
\label{Gl3}
\end{eqnarray}
In the latter equations have admitted for the possibility that 
${\widetilde m}$  
may not necessarily coincide in value with the magnetic 
quantum number $m$ in eq.~(\ref{Gl1}).
Our case is that eq.~(\ref{master}) is exactly solvable and although the
cotangent function ceases to commute with {\bf L}$^2$, the interaction retains
unaltered the level multiplicity characterizing the free rotator.
Nonetheless, the system is not rotationally invariant, 
the signature for this being the difference between
$F(\theta )$ from eq.~(\ref{Gl3}) and  $P^{|m|}_l(\theta )$ from
eq.~(\ref{Sch_pol}).

\section{The cotangent-hindered rigid rotator }

In order to solve eq.~(\ref{master})  we first change
variable to
\begin{equation}
F(\theta )=\frac{U(\theta )}{\sqrt{\sin \theta }}.
\label{Gl4}
\end{equation}
Next we observe that substitution of eq.~(\ref{Gl4}) into 
eq.~(\ref{master}) and subsequent
multiplication of both sides by $\sqrt{\sin \theta}$ amounts to
the following one-dimensional Schr\"odinger equation
(in dimensionless units) with the 
trigonometric Rosen-Morse potential, $V_{\mbox{RM}}(\theta)$, 

\begin{eqnarray}
\frac{{\mathrm d}^2U(\theta )}{{\mathrm d}\theta^2}
-V_{\mbox{RM}}(\theta)U(\theta )  +(\epsilon +\frac{1}{4} )U(\theta) &=&0,
\nonumber\\
V_{\mbox{RM}}(\theta )= \frac{ \overline{m}(\overline{m}+1)}
{\sin^2\theta }
-2b \cot \theta ,
&\quad&  \overline{m}=|{\widetilde m}|-\frac{1}{2}\geq 0.
\label{Gl5}
\end{eqnarray}
The trigonometric Rosen-Morse potential is among the well known
exactly solvable potentials managed by the super symmetric quantum mechanics
(SUSY-QM) and listed in \cite{Dutt}.
 We have studied  eq.~(\ref{Gl5}) earlier in ref.~\cite{CK_05}
and solved it independently in terms of the real
non-classical Romanovski polynomials \cite{Routh-Rom}
(reviewed in ref.~\cite{raposo})
instead of the Jacobi polynomials of purely imaginary arguments and 
complex conjugate parameters preferred by \cite{Dutt}.  
The spectrum of this potential problem is well known and given by
\cite{CK_05}, 

\begin{eqnarray}
\epsilon_t &=&-\frac{b^2}{
t(t+1) +\frac{1}{4}} 
+ t(t+1), \quad t:\stackrel{\mbox{def}}{=}n+| {\widetilde m}|=0,1,2,...,
\quad |{\widetilde m}|
=0,1,... t.
\label{qnmbs}
\end{eqnarray}
Observing that $t$ 
can take any non-negative integer value, and that 
$|{\widetilde m}|=0,1,... t$,  we conclude that the $t$ and $|{\widetilde m}|$ 
quantum numbers obey same branching rules as angular momentum and 
magnetic quantum number. 
In terms of these quantum numbers, the expressions for  
the wave functions are found as,
\begin{eqnarray}
F(\theta )\longrightarrow F_t^{|{\widetilde m}|}(\theta), 
\quad U(\theta)\longrightarrow U_t^{|{\widetilde m}|}(\theta),\quad
U_t^{|{\widetilde m}|}(\theta ) &=&N_{t|{\widetilde m}|}\sin ^{t+\frac{1}{2}}\theta
e^{-\frac{b\theta }{t+\frac{1}{2}}}
R_n^{  \frac{2b}{t+\frac{1}{2}}, -(t-\frac{1}{2}) }(\cot \theta ),
\label{Gl6}
\end{eqnarray}
where $R_n^{\alpha, \beta }(\cot \theta)$ denote the Romanovski polynomials,
and   we have  attached the 
indices $t$ and ${\widetilde m}$ to $F(\theta)$ and $U(\theta)$.
The Romanovski polynomials (reviewed in ref.~\cite{raposo})
satisfy the following differential hyper- geometric equation:
\begin{equation}
(1+x^2)\frac{{\mathrm d}^2R_n^{\alpha, \beta}}{{\mathrm d} x^2}
+2\left(\frac{\alpha }{2} +\beta x
\right)\frac{{\mathrm d}R_n^{\alpha, \beta}}{{\mathrm d}x}
-n(2\beta +n-1)R_n^{\alpha, \beta}=0.
\end{equation}
They are obtained from the following  weight function,
\begin{equation}
\omega ^{\alpha, \beta}(x)=(1+x^2)^{\beta -1}\exp(-\alpha \cot^{-1}x),
\end{equation}
by means of the the Rodrigues formula:

\begin{equation}
R^{\alpha, \beta}_n(x)=\frac{1}{\omega^{\alpha, \beta }(x)}
\frac{{\mathrm d}^n}{{\mathrm d}x^n}
\left[ (1+x^2)^n\omega^{\alpha, \beta}(x)\right].
\end{equation}
The parameters $\alpha $ and $\beta$ for the case under investigation are 
calculated as,
\begin{eqnarray}
\alpha =\frac{2b}{n+|{\widetilde m}|+\frac{1}{2}}, &&
\beta =-(n+|{\widetilde m}|+\frac{1}{2}) +1.
\label{param}
\end{eqnarray}

The labeling of the wave functions $U^{|m|}_n(\theta)$ corresponds to,
\begin{equation}
U_t^{|{\widetilde m}|}(\theta )=
N_{t|{\widetilde m}|}e^{-\frac{\alpha \theta}{2}}(1+\cot^2\theta)
^{-\frac{1-\beta }{2}}R_{n=t-|{\widetilde m}|}^{\alpha, \beta}(\cot \theta),
\end{equation}
with $\alpha$ and $\beta$ from eq.~(\ref{param}).

The functions 
\begin{equation}
{\mathcal X}_t^{{\widetilde m}}(\theta ,\varphi)=
F_t^{|{\widetilde m}|}(\theta )
e^{i{\widetilde m}\varphi}=\frac{U_t^{|{\widetilde m}|}(\theta )}
{\sqrt{\sin\theta}}e^{i{\widetilde m} \varphi},
\label{I4}
\end{equation}
determine the solutions of the cotangent-perturbed rigid rotator on $S^2$.

\begin{figure}
\resizebox{0.80\textwidth}{7.5cm}
{\includegraphics{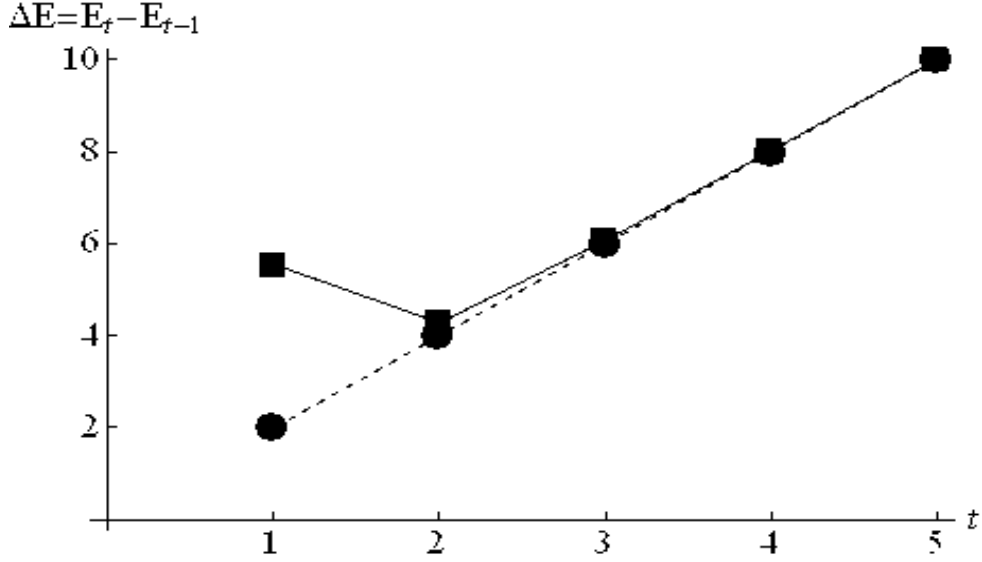}}
\caption{The splitting, $\Delta E=(E_{t}-E_{t-1})$,  for $b=1$,
between neighboring rotational levels belonging to
the ordinary rigid rotator (circles connected by a dashed line)
and the cotangent-perturbed rotator (squares connected by a solid line). 
Noticeable,  
the anomalous splitting between the ground state and its first excitation.
For the higher lying excitations  the perturbed level splittings 
rapidly approach the unperturbed ones, a behavior visualized by 
practically coincident squares and circles.
\label{rot_bds}}
\end{figure}
In summary:
\begin{itemize}
\item The branching rules between the non-negative
integers $t$ and $|{\widetilde m}|$ 
quantum labels introduced in eq.~(\ref{qnmbs}) are indistinguishable from
those of orbital  angular momentum and magnetic quantum number. 
In this sense the  energy spectrum, $\epsilon_t$,
in eq.~(\ref{qnmbs}) formally shows  same $(2t+1)$ fold
level multiplicity  as the unperturbed rotator.
The splitting between two subsequent levels reads,
\begin{equation}
E_t-E_{t-1}= 2t +\frac{2tb^2}{\left(t^2-\frac{1}{4}\right)^2}
\stackrel{t\to \infty }{\longrightarrow} 2t+{\mathcal O}
\left(\frac{1}{t^3}\right).
\label{main}
\end{equation}
Equation (\ref{main}) shows that for $b\sim 1$   
the perturbation affects mainly the  splitting between the ground state
and its first excitation and becomes rapidly negligible afterward.
In this fashion the rigid rotator hindered by a weak cotangent interaction
describes a rotational band with an anomalously large gap between the 
ground state and the first excited level according to 
$E_1-E_0=2+\frac{\mathbf 32}{\mathbf 9}b^2$
(see Fig.~1).

\item Though the  ${\mathcal X}^{m}_t(\theta, \varphi)$ functions
in eq.~(\ref{I4})
do not transform according to  a irreducible representation of 
${\mathbf L}^2$ in eq.~(\ref{I2}), for large $t$ values they 
asymptotically approach the spherical harmonics according to:

\begin{equation}
\sin^{t}\theta R_n^{  \frac{2b}{t+\frac{1}{2}}, 
-\left(t-\frac{1}{2}\right) }(\cot \theta )
\stackrel{t\to \infty }{\longrightarrow}
\sin ^{t }\theta R_{n=t-|m|}^{0, -\left(t-\frac{1}{2}\right)}(\cot \theta) 
\sim P_t^{|m|}(\cos \theta).
\end{equation}
This is due to the rapid flattening of the exponential (damping) factor
with the increase of $t$ (Fig.~2).

The last equation in the chain has already be
found in ref.~\cite{David} from a different perspective.
\end{itemize}

\begin{figure}
\resizebox{0.80\textwidth}{6.5cm}
{\includegraphics{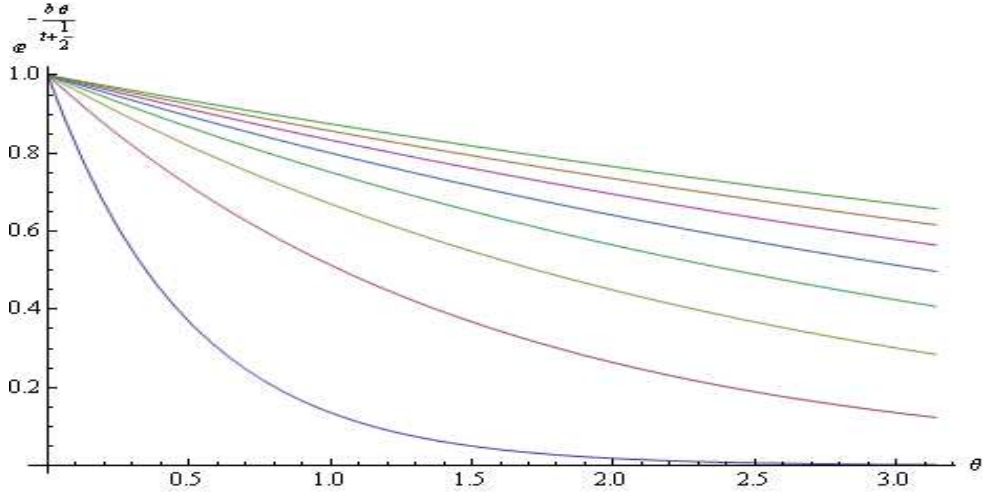}}
\caption{The dependence of the  damping factor (i.e. the exponential)
in eq.~(\ref{Gl6}) on $t$.  
The flattening with the increase of $t$ (from bottom to top) is well
pronounced.
\label{damping}}
\end{figure}
It is quite insightful to take a closer look on the structure of the
$F_t^{|{\widetilde m}|}(\theta)=
U^{|{\widetilde m}|}_t(\theta )/\sqrt{\sin \theta }$ functions in 
eq.~(\ref{Gl6}). Below we decompose  some of  the lowest
functions into associated Legendre functions. In omitting the 
normalization factor for simplicity, we find:

\begin{eqnarray}
n=1:&& {\widetilde m}=0:\quad  
F_1^0(\theta )=e^{-\frac{2b\theta }{3}}\sin \theta R_1^{\frac{4b}{3},
-\frac{1}{2}} (\cot \theta) =e^{-\frac{2b\theta }{3}}\left(
-P_1^0(\cos\theta ) +\frac{4b}{3}P_1^{1}(\cos\theta)\right),
\label{C1}\\
n=1:&& {\widetilde m}=1:\quad F_1^{1}(\theta )=
e^{-\frac{2b\theta }{3}}\sin \theta 
R_0^{\frac{4b}{3},-\frac{1}{2}} (\cot \theta)=
e^{-\frac{2b\theta }{3}}P_1^1(\cos\theta),
\label{C2}\\
n=2:&& {\widetilde m}=0:\quad 
F_2^0(\theta)= e^{-\frac{2b\theta }{5}}\sin^2\theta R_2^{\frac{4b}{5}, 
-\frac{5}{2}}(\cot \theta) =
e^{-\frac{2b\theta }{5}}{\Big(}\frac{1}{2}P_2^0(\cos \theta) +
\frac{8b}{15}
(P_2^1(\cos \theta)
\nonumber\\
& +&\frac{b}{5}P_2^2(\cos \theta)){\Big)},
\label{C3}\\
n=1:&& {\widetilde m}=1:\quad F_2^1(\theta)=
e^{-\frac{2b\theta }{5}}\sin^2\theta R_1^{\frac{4b}{5}, 
-\frac{5}{2}}(\cot \theta) =e^{-\frac{2b\theta }{5}}
\left( P_2^1(\cos \theta )-\frac{4b}{15}P_2^2(\cos \theta) \right),
\label{C}\\
n=0:&&{\widetilde m}=2:\quad 
F_2^2(\theta)=e^{-\frac{2b\theta }{5}}\sin^2\theta R_0^{\frac{4b}{5}, 
-\frac{5}{2}}(\cot \theta)=e^{-\frac{2b\theta }{5}}P_2^2(\cos \theta).
\label{C5}
\end{eqnarray}
The latter equations show that the wave functions 
$F_t^{|{\widetilde m}|}(\theta)$ of the cotangent-hindered rotator
are linear combinations of  damped 
associated Legendre functions, $P_l^{|m|}(\cos \theta )$, with
$l=t$ and ${\widetilde m}\in \lbrack {\widetilde m}, t\rbrack $. 
The transformation of the wave functions
of the perturbed problem to the wave functions of the free rotator is clearly
non-unitary and makes the breaking of the rotational symmetry at 
the level of the representation functions manifest. 
The above considerations provide a broader view on level degeneracy 
compared to the algebraic methods \cite{Iachello}, 
which conjecture that the symmetry of the Hamiltonian
is same as  the symmetry giving rise to
the observed degeneracies in the spectrum. Specifically,
the algebraic Hamiltonian 
of any type of a $(2l+1)$- fold level multiplicity
is supposed to be  described by means of a Hamiltonian that is a 
power series of
${\mathbf L}^2$. Specifically the algebraic Hamiltonian, ${\mathcal H}$,
for  the level multiplicity in eq.~(\ref{Gl6})
would be (in dimensionless units):
\begin{eqnarray}
{\mathcal H}-\frac{1}{4}&=&{\mathbf L}^2 -
\frac{b^2}{{\mathbf L}^{2}+\frac{1}{4}},
\label{alg_H}
\end{eqnarray}
effectively equivalent to
\begin{eqnarray}
{\mathcal H}-\frac{1}{4}&=&
{\mathbf L}^2 -\frac{b^2}{l^2(l+1)^2}
\frac{1}{\left[1+\frac{1}{4(l(l+1)}\right]}
{\mathbf L}^2,
\end{eqnarray}
in which case the standard spherical harmonics would remain eigenfunctions 
of the perturbed rotator,
\begin{eqnarray}
\left({\mathcal H}-\frac{1}{4}\right)
Y_l^m(\theta, \varphi)& =& \left[l(l+1)-\frac{b^2}{l(l+1) +\frac{1}{4}}\right]
Y_l^m(\theta, \varphi).
\label{I5}
\end{eqnarray}

The difference between the algebraic method in eqs.~(\ref{alg_H})-(\ref{I5}),
on the one side, and the 
Hamiltonian method in eqs.~(\ref{Gl5}), (\ref{Gl6}), and (\ref{I4})
will become detectable  at the level of transition probabilities.

\section{Conclusions}
\begin{figure}
{\includegraphics[width=7.5cm]{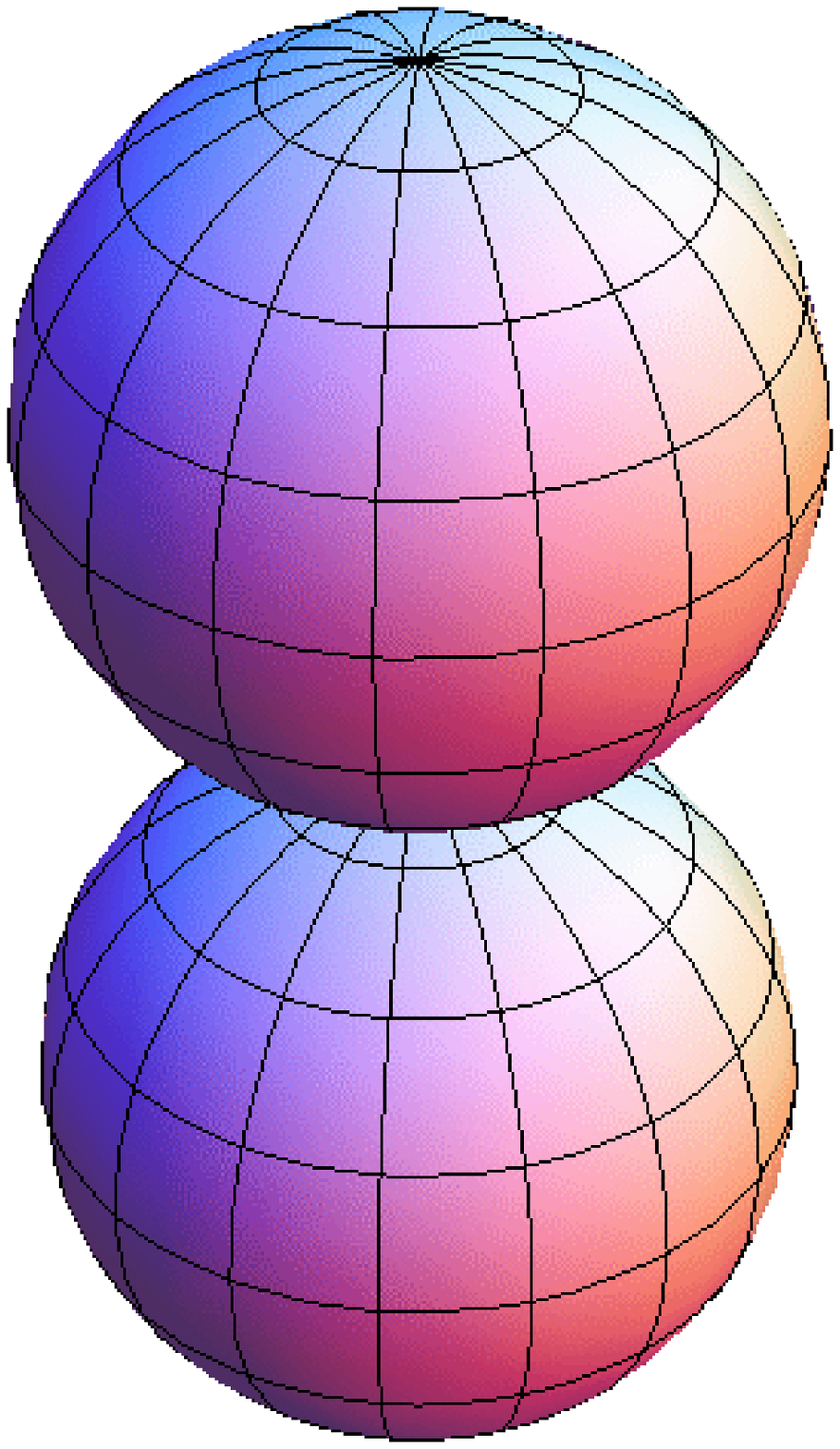}}
{\includegraphics[width=8.5cm]{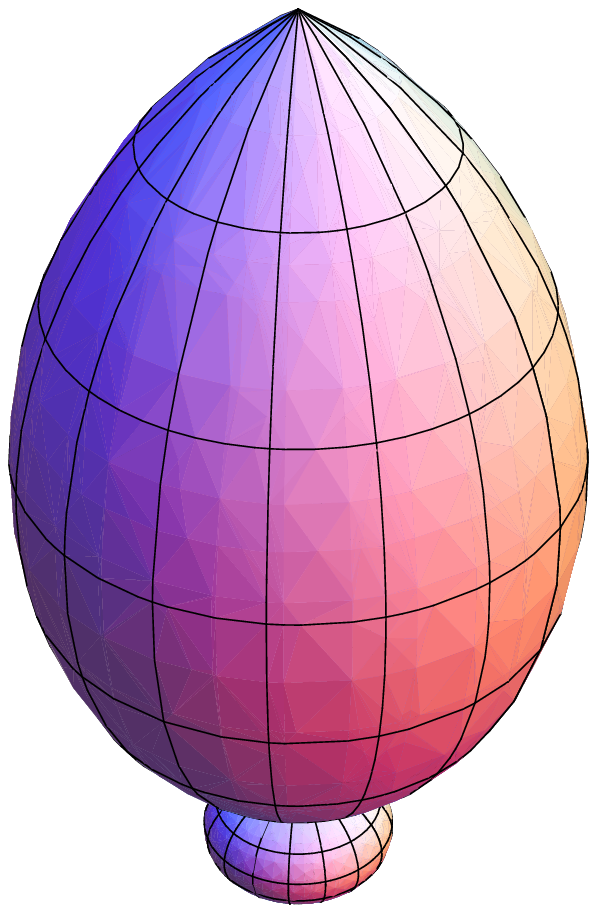}}
\caption{
The regular spherical harmonic $|Y_1^0(\theta , \varphi)|$ (left)
versus the damped, $|{\widetilde Y}_1^0(\theta, \varphi)|$, (right).
\label{tld1}}
\end{figure}

\begin{figure}
{\includegraphics[width=7.5cm]{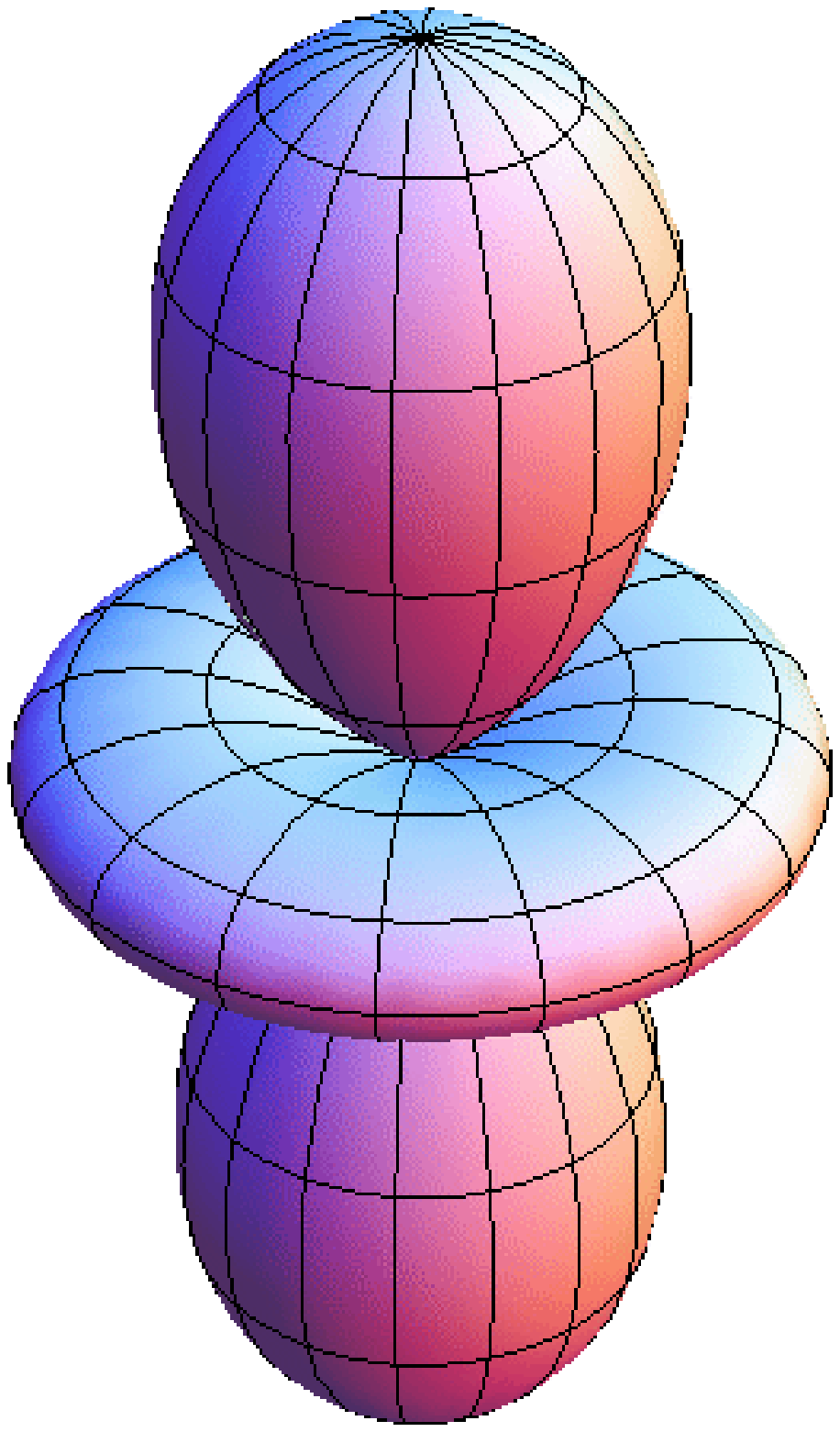}}
{\includegraphics[width=6.5cm]{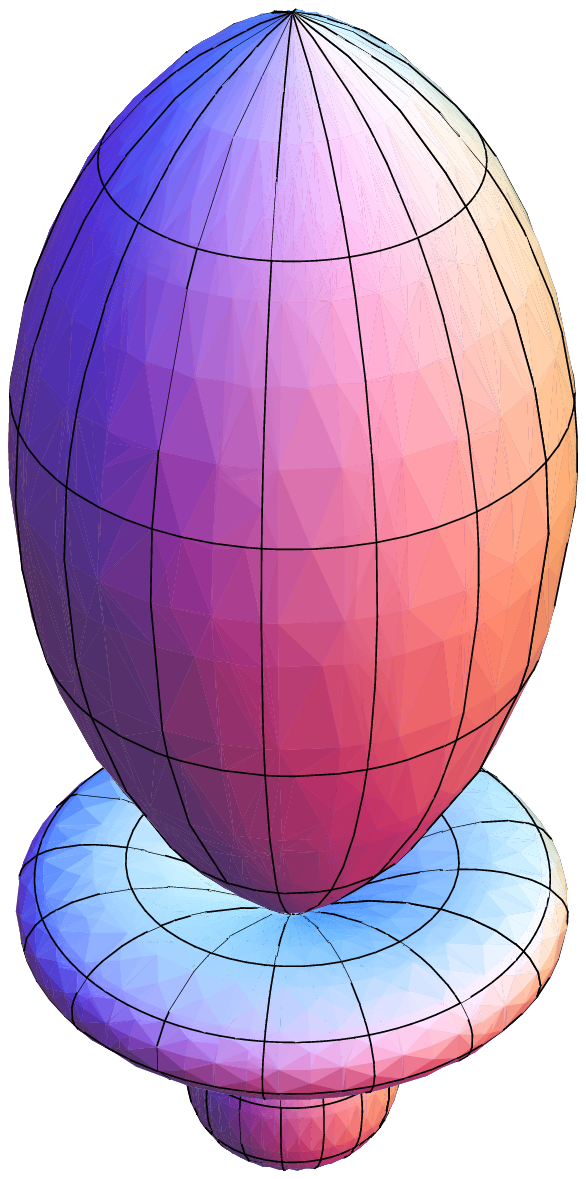}}
\caption{
The regular spherical harmonic $|Y_2^0(\theta , \varphi)|$ (left)
versus the damped, $|{\widetilde Y}_2^0(\theta, \varphi)|$, (right).
\label{tld2}}
\end{figure}

In this work we considered the option of having a symmetry
breaking at the level of the representation functions while preserving the 
degeneracy patterns in the spectra of the original symmetry. 
It this fashion, we argued that
a symmetry breaking opaqued by level degeneracy can take place.
On the example of a cotangent-perturbed rigid rotator we showed that 
the $(2l+1)$-fold level degeneracy is a softer criterion for 
rotational symmetry realization, 
the rigorous one being the property of the wave functions to transform
according to  irreducible representations of {\bf L}$^2$ in (\ref{I2}).
In eqs.~(\ref{C1})--(\ref{C5}) we noticed that the  solutions,
$F(\theta )$, of eq.~(\ref{master}) can be represented as linear 
combinations of (exponentially) damped associated Legendre functions.
In terms of the latter,  damped   spherical harmonics, 
${\widetilde Y}_t^{ m}(\theta ,\varphi)$, 
can be defined  according to,
\begin{eqnarray}
e^{-\frac{\alpha \theta}{2}}{\mathbf L}^2 e^{\frac{\alpha \theta }{2}}
{\widetilde Y}_t^{ m}(\theta, \varphi)&=&
l(l+1){\widetilde Y}_t^{ m}(\theta, \varphi), \nonumber\\
{\widetilde Y}_t^{ m}(\theta, \varphi)&=&
e^{-\frac{\alpha \theta}{2}}P_t^m(\theta) e^{im\varphi}
\label{Y_tilde}
\end{eqnarray}
In Figs.~\ref{tld1}-\ref{tld2} we compare the 
regular and damped spherical harmonics for illustrative purposes.

The similarity transformation of ${\mathbf L}^2$
in the last equation is a dilation transformation which shows that 
the free and the hindered rotator solutions are related by a non-unitary 
transformation. The equations (\ref{C1})--(\ref{C5}) in combination with
 (\ref{Y_tilde}) allow to give the cotangent-hindered
rotator also  the reading of a 
realization of an so(3) algebra unitarily inequivalent to the standard
rotational. 
Specifically for the case of a weak perturbance,
we revealed possibility of observing in diatomic 
molecules rotational bands 
characterized by an anomalously large
gap between the ground state of the rigid rotator
and its first excitation, while practically preserving the subsequent 
level splittings.

\section{Acknowledgments}

Work partly supported by CONACyT-M\'{e}xico under grant number
CB-2006-01/61286.

\end{document}